\documentclass[dvips]{article}
\usepackage{icrctc07}

\title{VEGAS, the VERITAS Gamma-ray Analysis Suite}
\shorttitle{VEGAS}
\authors{P. Cogan$^1$ for the VERITAS Collaboration$^2$.}
\shortauthors{P. Cogan and et al}
\afiliations{$^1$McGill University, 3600 University Street, Montreal, QC H3A 2T8, Canada \\ 
$^2$For full author list see G. Maier, "Status and Performance of VERITAS", these proceedings}
\email{coganp@hep.physics.mcgill.ca}

\abstract{VERITAS, the Very Energetic Radiation Imaging Telescope Array System, is an array of four 12 m diameter imaging atmospheric Cherenkov telescopes for gamma-ray astronomy above 100 GeV currently in operation in Arizona. The VERITAS Collaboration has developed VEGAS, the VERITAS Gamma-ray Analysis Suite, a data-analysis software package for the processing of single- and multiple-telescope data produced by the array. The package consists of a core of six stages as well as visualisation and diagnostic components. It has been developed in C++ using modern objected-oriented design patterns to be highly flexible, configurable and extendable. VEGAS utilises CERN's ROOT data-analysis framework and runs on Linux and Mac OS X systems. The architecture and structure of the VEGAS package will be described in detail while the data analysis algorithms are described in additional papers.}

\begin{document}
\maketitle
%Begin the section.

\section{Introduction}

The VERITAS \cite{maier07a} Collaboration instituted a working group
in late 2005 to design, create and maintain a complete offline
analysis package for the reduction of VERITAS data. This group, named
the Offline Analysis Working Group (OAWG), comprises collaborators
from several of the VERITAS institutions. Following a series of
meetings at various institutions in Ireland, Canada and the U.S., the
OAWG released the first version of the package VEGAS in early
2007. The OAWG has continued to release upgrades and expanded versions
of the package on a regular basis. This paper gives an overall summary
of the many components of VEGAS including analysis stages,
architecture, documentation, organisation and performance.

\section{Analysis Stages}

There are six stages in the VEGAS analysis paradigm as outlined in
Table \ref{vegas}. In the first stage calibration parameters such as
pedestals and relative gains are calculated. The VERITAS database
(implemented in MySQL\cite{mysql}) is queried for run-specific
information such as high voltage records, tracking information and
target information. These data are stored in a customised binary
output format generated using the CERN ROOT \cite{root} library.

In the second stage of the analysis the FADC traces are
evaluated\cite{cogan07}. This includes pedestal subtraction and
relative gain and timing correction. For each data channel in each
event, a data class containing the integrated charge, integration
parameters and other channel parameters are stored. The data are saved
to the same binary file.

In the third analysis stage, the data are cleaned \cite{daniel07}
using an advanced form of the Picture/Boundary cleaning
method. Following this the data are parameterised for both a 1-D and a
2-D analysis according to the Hillas \cite{hillas85} moment technique.

In the fourth analysis stage, stereo reconstruction is applied (if the
data involves more than one telescope). Cuts are applied to determine
whether the image from a particular telescope is of sufficient quality
to contribute to the stereo parameterisation. Shower origin in the sky
and shower core location on the ground are determined. These are used
in conjunction with Monte Carlo simulations of gamma-ray air showers
to calculate mean scaled-width, mean scaled-length and an energy for
each event \cite{daniel07}.

In the fifth analysis stage, background rejection is performed. This
is done primarily using cuts on mean-scaled width and length, although
cuts on other parameters such as time can also be applied. Stage 5 has
the ability to cut on multiple data types including single telescope
Hillas parameters, stereo parameters or a combination of both.

The sixth analysis stage performs background estimation in various
observing modes for both 1-D and 2-D analysis. The
reflected-region background model and ring-background models
\cite{daniel07} have been implemented. The probability of detection of
an observed target is calculated using the Li\&Ma \cite{lima83}
formalism, and where a sufficiently strong detection is made, a
spectrum can be determined.

\begin{table*}[htbp]
  \begin{center}
 \begin{tabular}{ |c|c|c|c|c|c| }
   \hline
Stage&Purpose&Input(s)&Output&Time(m)&Size(MB)\\
\hline
1 & Calib. Calculation & Raw Data              & Calibration Data      & 8.2 & 51   \\ \hline
2 & Calib. Application & Raw + Calibration Data& Calibrated Events     & 39  & 6200 \\ \hline
3 & Image Param.       & Calibrated Events     & Param'd Events        & 14  & 224  \\ \hline
4 & Shower Recon       & Param'd Events        & Recon'd Showers       &  6  &  40  \\ \hline
5 & Event Selection    & Recon'd Showers       & Selected Events       & $<2$& 202  \\ \hline
6 & Results            & Selected Events       & Statistics \& Figures & $<1$& $<1$ \\ \hline
  \end{tabular}
  \end{center}
  \caption{The VEGAS analysis chain with execution time (on a
  $2.66\:\mathrm{GHz}$ Apple XServe running OS X 10.4.9) and output
  size (for a 5000MB input file). Note that `Calib.', `Param'd' and
  `Recon'd' are abbreviations of `Calibration', `Parameterised' and
  `Reconstructed' respectively.}
\label{vegas}
\end{table*}

\section{Architecture}

Modularity is one of the primary VEGAS design goals. This means that for any given algorithm, for which there could be several implementations, it must be simple to substitute one implementation for another. This is accomplished using the object-oriented language C++ with \emph{singleton} instances of \emph{factory} classes. A factory class selects the algorithm that is to be used based on the configuration specified, and the singleton pattern ensures that only one factory class is generated for each algorithm. The use of factory classes means that function calls do not need to check which algorithm is being used each time the function is called. This leads to reduced analysis overhead and simpler code.

\section{Configuration}

Each analysis class can be configured from the command line or from a
configuration file (or both). The configurable options for each class
are set when the class is constructed. Users can easily get a list of
the configurable options available from the command line help. However, as
all of the main programs have a very large number of configurable
options, it is often desirable to use a configuration file. Each main
program can automatically produce a configuration file indicating the
default values of every configurable parameter. The user can then edit
this configuration file and use it as a pilot file for the main
program. Users can also easily share configuration files, which in
concert with the use of tagged releases of VEGAS, ensures that
reproducible results can always be attained. The configuration is also
written to the data file allowing the settings with which the analysis
was run to be queried.

\section{Data Products}

All of the VEGAS data classes are constructed as ROOT TObjects,
allowing reading and writing in an efficient manner with a standard
interface. Within the ROOT file, output from each analysis stage is
arranged into directories, with some output saved as single instances
of the corresponding data class, and some output saved as ROOT
TTrees. The TTree can store multiple instances of a data class,
allowing large volumes of data to be read and written in an efficient
manner.

\section{Macros and Visualisation}

\begin{figure}
\begin{center}
\includegraphics [width=0.48\textwidth]{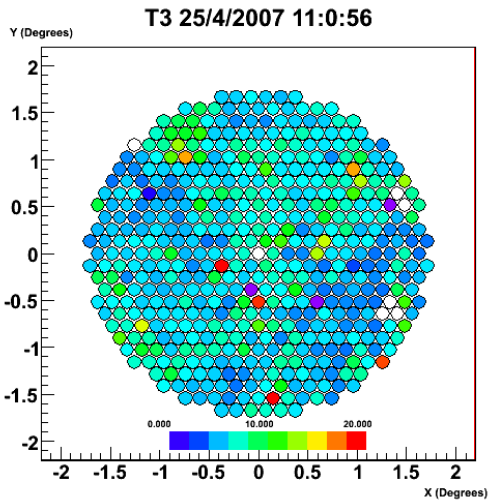}
\end{center}
\caption{Macros used in the ROOT CINT command line interpreter in
conjunction with the VEGAS shared library can be used to easily
visualise data such as the PMT currents at a particular
time.}\label{macro}
\end{figure}

The VEGAS data class definitons are compiled into a shared library
which can be used to develop simple macros which run within the ROOT
CINT interpreter. The output of such a macro is displayed in Figure
\ref{macro}, showing the PMT currents in each pixel at a particular
time for a single telescope.

A diagnostics program is also built using the ROOT library which
displays reconstructed events in the camera, shower and mirror
planes. This diagnostics program also displays the raw FADC data and
demonstrates image cleaning, Hillas parameterisation,
and stereo reconstruction. There are also built-in macros for
examining Monte Carlo simulations of air showers.

\section{Simulations}

Monte Carlo simulations \cite{maier07b} of gamma-ray and comic-ray air
showers play a pivotal role in the analysis of Cherenkov telescope
data. There are a variety of simulation channels available within the
VERITAS collaboration including KASCADE \cite{kascade}, Corsika
\cite{corsika}, GrISU \cite{grisu} and ChiLA. The reading and writing
of simulation data specific to each package is handled using
package-specific data classes. Although simulations for VEGAS have
primarily been performed using KASCADE, VEGAS is designed to be
compatible with the other Monte Carlo codes so that systematic bias
may be avoided, and simulations codes can be cross-checked.

\section{Documentation}

The VEGAS code is documented in several ways. At the lowest level,
functions and classes are documented using Doxygen \cite{doxygen}. Doxygen allows
code and comments to be viewed in a hyperlinked format using a
standard web browser. The OAWG developers can expand on the
documentation, particularly in relation to algorithms, using a
customised Wiki \cite{wiki}. The Wiki is a website that easily allows fast content
editing - this is beneficial in the case where algorithms and methods
are being continuously updated. Is it also used to coordinate meetings
and consistency checks. Finally, an online users manual is maintained
at a separate website. The online manual details compilation and
installation of dependencies and the code itself. Detailed
instructions regarding running the code and interpreting the output are
also provided.

\section{Organisation and Communication}

The OAWG is organised through weekly teleconferences. This affords
both developers and users the opportunity to discuss and query issues
relevant to algorithms and analysis techniques. This is
further facilitated by the use of two Electronic Logbooks (ELog)
\cite{elog}. The first is the main OAWG ELog ; it is a forum where
developers can discuss and debate various analysis and coding
issues. This has a proven to be a valuable resource as developers can
use it to keep track of changes while maintaining a long term archive
of discussions pertaining to analysis and code design. The second ELog
is a users forum which facilitates VEGAS users in asking questions
regarding how to install and run VEGAS.

\section{Directory Structure}

Each main analysis stage has its own directory, containing class
definitions, implementations, documentation and examples. There is a
\emph{common} directory containing classes and algorithms that are
used by multiple analysis stages. These include most data classes,
configuration classes and classes pertaining to I/O. There are also
separate directories for macros, utilities, diagnostics, documentation
and binaries.

\section{Code Management}

The VEGAS source code is stored in the main VERITAS CVS \cite{cvs}
repository. CVS is a version control system which allows several
developers to concurrently work on the same code. It also allows old
versions of the code to be easily retrieved. The CVS repository can be
browsed in a hyperlined envrionment \cite{viewcvs} facilitating the
examination of code revisions. Code tagging is used to enable
users and developers to download identical versions of the code for
testing. This ensures reproducability when users are cross-checking
results.

In order to ensure consistency and stability, the VEGAS code is
released following beta-test cycles. When new analysis or coding
features have matured, the current version of VEGAS on CVS is
tagged. This tagged code undergoes a series of beta tests and cross
checks in order to ensure stability. One of the most important tests
is ensuring that the code gives identical output on different
platforms and architectures and with different compiler versions. Once
a tag has been approved it is released to the VERITAS collaboration
for general use. In the event that coding errors or other issues are
found in the released version, it can be corrected as a separate CVS
branch. This allows development of the current version to continue
without delaying deployment of the corrected release.

\section{Performance}

VERITAS raw data is particularly large with approximately
$5\:\mathrm{GB}$ of data written for a typical run. These runs
contain approximately 240,000 to 260,000 events. The analysis times
and output file sizes for each analysis stage (for a single 5GB run) are shown
in Table \ref{vegas}. As the output of stage two is particularly
large, it can be discarded after running stage three resulting in more
manageable file sizes.

\section{Future Plans}

Although a huge amount of development has already taken place, there
remains much to be done. The development of VEGAS will continue
hand-in-hand with the development of new algorithms for stereo event
reconstruction, background estimation, 2-dimensional analysis and
spectral analysis. 

\section{Acknowledgments}

This research is supported by grants from the U.S. Department of
Energy, the U.S. National Science Foundation, the Smithsonian
Institution, by NSERC in Canada by PPARC in the UK and by Science
Foundation Ireland.

%This is the reference to .bib file (Whitout .bib!)
\bibliography{icrc0652}
%This in the bibtex style, is ok.
\bibliographystyle{plain}

\end{document}